\documentclass[conference]{IEEEtran}
\IEEEoverridecommandlockouts
\usepackage{cite}
\usepackage{amsmath,amssymb,amsfonts}
\usepackage{algorithmic}
\usepackage{graphicx}
\usepackage{booktabs} 
\usepackage{siunitx}
\usepackage{multirow}
\usepackage{multicol}
\usepackage{comment}
\usepackage{caption} 
\usepackage{textcomp}
\usepackage{xcolor}
\def\BibTeX{{\rm B\kern-.05em{\sc i\kern-.025em b}\kern-.08em
    T\kern-.1667em\lower.7ex\hbox{E}\kern-.125emX}}
\begin{document}

\title{Learning Frame Similarity using Siamese networks for Audio-to-Score Alignment\\
\thanks{This project has received funding from the European Research Council (ERC) under the European Union's Horizon 2020 research and innovation programme under the Marie Skłodowska-Curie grant agreement No. 765068.}
}

\author{\IEEEauthorblockN{Ruchit Agrawal}
\IEEEauthorblockA{\textit{Centre for Digital Music} \\
\textit{Queen Mary University of London}\\
London, United Kingdom \\
r.r.agrawal@qmul.ac.uk}
\and
\IEEEauthorblockN{Simon Dixon}
\IEEEauthorblockA{\textit{Centre for Digital Music} \\
\textit{Queen Mary University of London}\\
London, United Kingdom \\
s.e.dixon@qmul.ac.uk}
}

\maketitle

\begin{abstract}
Audio-to-score alignment aims at generating an accurate mapping between a performance audio and the score of a given piece. Standard alignment methods are based on Dynamic Time Warping (DTW) and employ handcrafted features, which cannot be adapted to different acoustic conditions.
We propose a method to overcome this limitation using learned frame similarity for audio-to-score alignment. We focus on offline audio-to-score alignment of piano music.  
Experiments on music data from different acoustic conditions demonstrate that our method achieves higher alignment accuracy than a standard DTW-based method that uses handcrafted features, and generates robust alignments whilst being adaptable to different domains at the same time. 
\end{abstract}


\section{Introduction}
\label{introduction}
 The significance of neural networks for signal processing was pointed out early by \cite{hwang1997past, luo1999applied}, and their efficacy for Music Information Retrieval (MIR) has been demonstrated for a variety of tasks like 
music generation \cite{eck2002first},
music transcription \cite{hawthorne2018onsets} as well as music alignment \cite{dorfer2018learning}.
Audio-to-score alignment is the task of finding the optimal mapping between a performance and the score for a given piece of music. Dynamic Time Warping (DTW) \cite{sakoe1978dynamic} has been the de facto standard for this task, typically incorporating handcrafted features \cite{dixon2005line, ewert2009high, arzt2012adaptive}. The primary limitation of handcrafted features lies in their inability to adapt to different acoustic settings and thereby model real world data in a robust manner, in addition to not being optimized for the task at hand. 

\par This paper presents a novel method for DTW-based audio-to-score alignment, which does not depend on handcrafted features, but learns them directly from the music data at the frame level.
We propose learning a frame similarity matrix using neural networks which is then passed on to a DTW algorithm that computes the optimal warping path through the matrix, yielding the alignment. 
We propose the use of twin Siamese networks \cite{bromley1994signature} each containing a Convolutional Neural Network (CNN) \cite{lecun1999object} architecture for learning frame similarity. 
The advantage of our method is that it is
is efficiently able to learn meaningful representations for DTW directly from data and is thereby adaptable to different acoustic settings.

We conduct experiments on piano music using our approach and test its performance on the Mazurka dataset \cite{sapp2007comparative}, which contains recordings from different eras spanning various acoustic conditions;
and demonstrate improvements over \emph{MATCH} \cite{dixon2005match}, a standard DTW-based method that uses handcrafted features. We additionally explore two methods to improve the performance of our baseline models, namely salience representations \cite{bittner2017deep} and data augmentation. 
\par To the authors' knowledge, this is the first method to employ learned frame similarities using Siamese CNNs for audio-to-score alignment. Additionally, this is the first method to incorporate pitch salience for audio-to-score alignment to the authors' knowledge. The rest of the paper is organized as follows: We describe prior work and our relation to it in Section \ref{related}. Section \ref{method} details our proposed method and model pipeline. The experimentation conducted and results obtained using our method are described in Section \ref{experiments}. We present the conclusions of the present research and highlight  possible  directions  for  future  work in Section \ref{conclusion}.
\section{Related Work}
\label{related}
Early works on feature learning for 
Music Information Retrieval 
(MIR) employ algorithms like Conditional Random Fields \cite{joder2013learning} or deep belief networks \cite{schmidt2012feature}, whereas 
recent work in this direction is moving towards the usage of deep neural networks \cite{thickstun2016learning}.
\begin{figure*}[th]
  \centering
  {\includegraphics[width=6.5in]{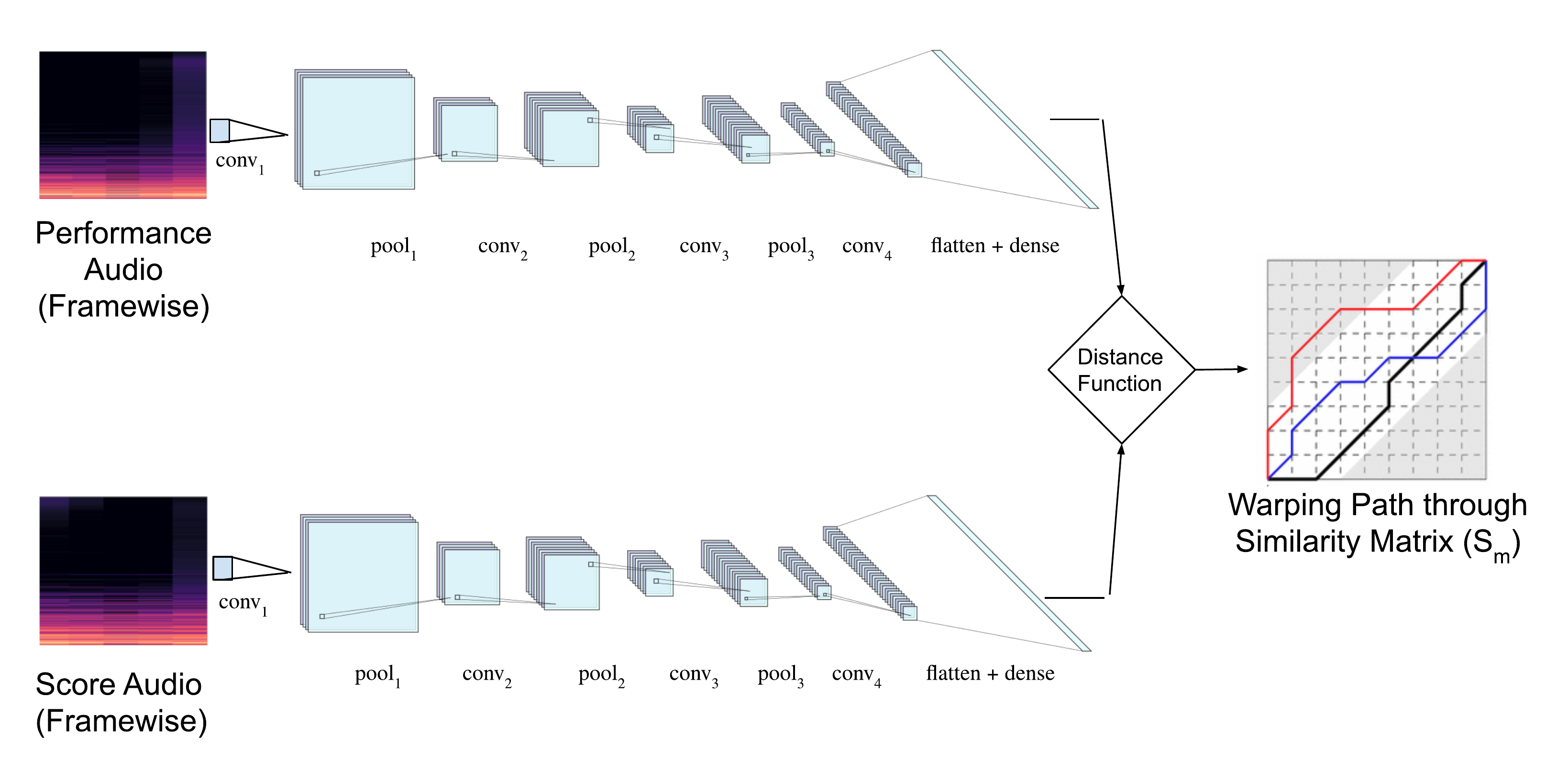}}
  \caption{Model Pipeline}
  
\begin{tabular}{r@{ : }l r@{ : }l}
$conv_i$ & \textit{$i_{th}$ convolution layer} & $pool_j$ & \textit{$j_{th}$ pooling layer}\\
$flatten$& \textit{Flatten layer} & $dense$& \textit{Fully connected layer} 
\end{tabular}
  \label{fig:pipeline}
  \vspace{-0.4cm}
\end{figure*}
Work specifically on learning features for audio-to-score alignment has focused on the evaluation of current feature representations \cite{joder2010comparative}, learning features for alignment using a Multi Layer Perceptron \cite{izmirli2010understanding},
and learning a mapping several common audio representations based on a best-fit criterion \cite{joder2011optimizing}. Recently, transposition-invariant features were proposed for music alignment \cite{arzt2018audio}, however these features while being robust to transposition, are sensitive to large tempo variations and underperform in such situations. 
\cite{dorfer2018learning2} is a recent work on score following, a task related to audio-to-score alignment.
While they employ reinforcement learning to train a score follower in real time, we focus on robust offline alignment across various acoustic conditions using frame similarity learning. 

\par Another direction which sets the context for our work is sound similarity; 
approaches to which include capturing music segment similarity using two-dimensional Fourier-magnitude coefficients \cite{nieto2014music}, similarity network fusion to combine different frame-level features for hierarchy identification of repeated sections in music \cite{tralie2019enhanced}, and application of Siamese Neural Networks for content-based audio retrieval \cite{manocha2018content}. The closest work to ours which employs the notion of learned sound similarity for music alignment is 
\cite{izmirli2010understanding}, to the authors' knowledge. While they use a Multi-Layer Perceptron 
to compute if two frames are the same or not, we compute frame similarity using Siamese CNNs. In addition to using an enhanced framework which is suitable for the similarity detection task, our work differs from them in that we also compute the extent of similarity in the form of non-binary distances and use this distance (or dissimilarity) matrix further for alignment. We additionally employ deep salience representations, which prove to be an effective method to improve alignment accuracy over our baseline models.
\section{Proposed Methodology}\label{method}
We propose a novel method for DTW-based audio-to-score alignment that uses Siamese neural networks. We additionally employ deep salience representations \cite{bittner2017deep} to improve model performance in data-scarce conditions. We describe the method in detail in the subsequent subsections.
\subsection{Siamese Convolutional Neural Networks}\label{siameseCNN}
 The standard feature representation choice for music alignment is a time-chroma representation \cite{bartsch2005audio} generated from the log-frequency spectrogram, which is not trainable on real data, and thereby not adaptable to different acoustic settings.  
We override the feature engineering step and focus on learning frame similarity using Convolutional Neural Networks (CNNs), since they can jointly optimize the representation of input data conditioned on the similarity measure being used.
We employ a Siamese Convolutional Neural Network, a class of neural network architectures that contains two or more identical subnetworks \cite{bromley1994signature} for this task. 


We train a Siamese CNN, akin to that prototyped in \cite{agrawal2020hybrid}, to compute a frame similarity matrix $S_m$ to be fed to DTW to generate alignment. Figure \ref{fig:pipeline} gives an overview of our model pipeline. 
In order to keep the modality constant, we first convert the MIDI files to audio through FluidSynth \cite{henningsson2011fluidsynth} using piano soundfonts. 
The two audio inputs are converted to a low-level spectral representation using a Short Time Fourier Transform, with a hop size of 23 ms and a hamming window of size 46 ms. 
Our training data contains synchronized audio and MIDI files, so it is straightforward to extract matching frame pairs. For each matching pair, we randomly select a non-matching pair (using MIDI-information) in order to have a balanced training set.
The inputs to the Siamese network are labelled frame pairs from the performance audio and the synthesized MIDI respectively.
We employ the contrastive loss function \cite{hadsell2006dimensionality} while training our models. We choose this formulation over a standard classification loss function like cross entropy since our objective is to differentiate between two audio frames. Let $X = (X_1, X_2)$ be the pair of inputs $X_1$ and $X_2$, $W$ be the set of parameters to be learnt and $Y$ be the target binary label ($Y$ = 0 if they match and 1 if otherwise). Task-specific loss functions have shown promising results in the fields of image processing and natural language processing \cite{qi2017contrastive, amirhossein2018multi}. The contrastive loss function 
for each tuple is computed as follows:
\begin{equation}
    L(W, X, Y) = (1-Y)\frac{1}{2}(D_W)^2 + (Y)\frac{1}{2}\{max(0, m - D_W)\}^2
\end{equation}
where $m$ is the margin for dissimilarity and $D_W$ is the Euclidean Distance between the outputs of the subnetworks. Pairs with dissimilarity greater than $m$ do not contribute to the loss function. 
More formally, $D_W$ can be expressed as follows:
\begin{equation}
    D_W(X) = \sqrt{\{G_W(X_1) - G_W(X_2)\}^2}
\end{equation}
where $G_W$ is the output of each twin subnetwork for the inputs $X_1$ and $X_2$. Since it is a distance-based loss, it tries to ensure that semantically similar examples are embedded close to each other, which is a desirable trait for extracting alignments. 
\par The Siamese network thus learns to classify the sample pairs as similar or dissimilar. 
This is done for each audio frame pair and the similarity matrix thus generated is then passed on to a 
DTW-based algorithm to generate the alignment path.
DTW generates an alignment between two sequences $A$= $(a_1,a_2,...,a_m)$ and $B$ = $(b_1,b_2,...,b_n)$ by comparing them using a local cost function, at each point, with the goal of minimizing the overall cost. The path which yields this minimum overall cost is then the optimal alignment between the two sequences. Formally, it can be represented as follows:

\begin{equation}
D(i, j)  = d(i, j) + min\begin{cases}
D(i, j-1) \\ D(i-1, j) \\  D(i-1,  j-1)
\end{cases}
\end{equation}
where $d(i, j)$ is the distance measure (local cost) between points $a_i$ and $b_j$; and $D(i, j)$ is the total cost for the path which generates the optimal alignment between the sequences $A_{1..i}$ and $B_{1..j}$. We employ Euclidean distance as our distance measure and the DTW framework from \cite{giorgino2009computing} to compute the warping paths. 
\subsection{Deep Salience Representations}
\begin{figure}[ht]
  \centering
  {\includegraphics[width=\columnwidth]{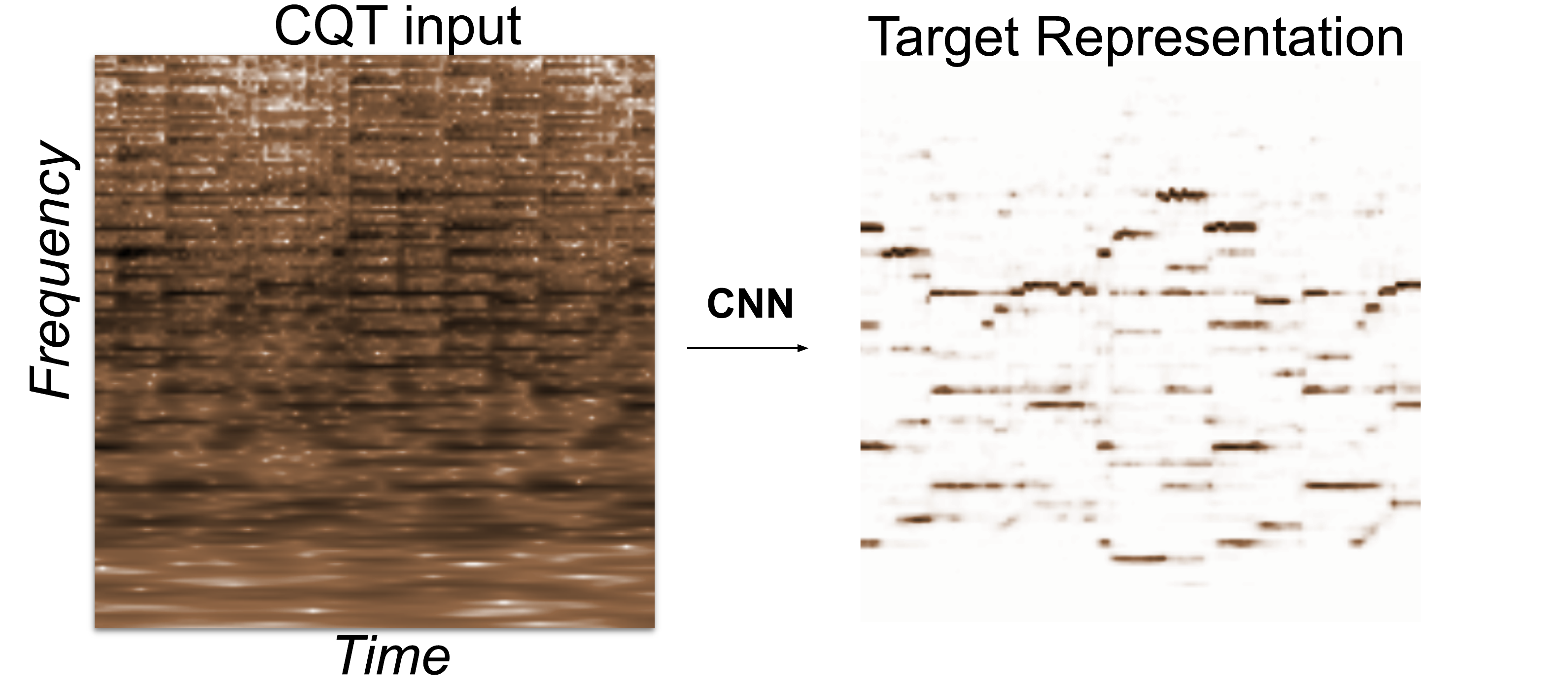}}
  \caption{Salience representations to address data sparsity}
  \label{fig:salience}
\end{figure}
We employ deep salience representations \cite{bittner2017deep} for effective training of our models. These are time-frequency representations aimed at estimating the likelihood of a pitch being present in the audio. Figure \ref{fig:salience} shows an example of a salience representation. The primary motivation behind using such a representation is that it de-emphasizes non-pitched content and emphasizes harmonic content, thereby aiding training in data-scarce conditions.
We employ the model proposed by \cite{bittner2017deep}, trained to learn a series of convolutional filters, constraining the target salience representation to have values between 0 and 1, with larger values corresponding to time-frequency bins where fundamental frequencies are present. The model is trained to minimize the cross entropy loss as follows:
\begin{equation}
    L(y, \hat{y}) = -y log(\hat{y}) - (1 - y)  log(1 - \hat{y})
\end{equation}
where both $y$ and $\hat{y}$ are continuous values between 0 and 1. 
\par We compare the performance obtained using salience representations with that obtained using the Short-Time Fourier Transform (STFT) and Constant-Q Transform (CQT) of the raw audios. We employ these input representations for comparative purposes. We employ a hop size of 23 ms and a hamming window of size 46 ms. We employ a CQT with 24 bins per octave, with the first bin corresponding to frequency 65.4 Hz (midi note C2).
\section{Experiments and Results}\label{experiments}
\subsection{Experimental Setup}\label{setup}
We employ the MAPS database \cite{emiya2009multipitch}, the Saarland database \cite{muller2011saarland} and the Mazurka dataset \cite{sapp2007comparative} for our experiments. From the original MAPS database, which contains synthesized MIDI-aligned audio for a range of acoustic settings, we select the subset \emph{MUS} containing complete pieces of piano music, and append it to the Saarland database. We split the resultant database comprising 288 recordings randomly into sets of 230 and 58 recordings. These sets form our training and validation sets respectively.
We test the performance of our models on the Mazurka dataset \cite{sapp2007comparative}, which contains recordings of Chopin's Mazurkas dating from 1902 to the early 2000s, thereby spanning across various acoustic settings. This dataset contains annotations of beat times for five Mazurka pieces. The alignment error for these pieces has a standard deviation of 11 ms.
 \begin{table}[ht]
   \caption{Architecture of our model}
   \centering
   \begin{tabular}{ c c c c c } \toprule
       \textbf{Type of layer} & \textbf{Input size} & \textbf{Kernels} & \textbf{Kernel size}    \\ \midrule
       Convolution  & $128 * 128 * 3$ & 64 &   $5 * 5$ \\ \midrule
       Max-Pooling & $128 * 128 * 64$  & 1 &  $2 * 2$   \\ \midrule
       Convolution  & $64 * 64 * 64$ & 128 & $5 * 5$ \\ \midrule
       Max-Pooling & $64 * 64 * 128 $ & 1 & $2 * 2$   \\ \midrule
       Convolution  & $32 * 32 * 128$ & 256  & $3 * 3$  \\ \midrule
       Max-Pooling & $32 * 32 * 256$ & 1 & $2 * 2$   \\ \midrule
       Convolution  & $16 * 16 * 256$ & 512 & $3 * 3$  \\ \midrule
       Flatten & $16 * 16 * 512$ & - & -  \\ \midrule
       Fully Connected  & $131072$  & - & - \\ \midrule
      \bottomrule
\end{tabular}
\label{tab:arch}
\end{table} 

\begin{table*}[th]
\caption{Results of our models}
   \centering
\begin{tabular}{ccccccccc} \toprule
\hline 
\multirow{2}{*}{\textbf{Model}} & \multicolumn{4}{c}{\textit{Binary Matrix}} & \multicolumn{4}{c}{\textit{Distance Matrix}}
\tabularnewline
  & \textbf{$<$25ms}& \textbf{$<$50ms} & \textbf{$<$100ms} & \textbf{$<$200ms} & \textbf{$<$25ms} &\textbf{$<$50ms} & \textbf{$<$100ms} & \textbf{$<$200ms}\\
\midrule 
 $MATCH$\cite{dixon2005match} & - & - & - & -  & 64.8 & 72.1 & 77.6 & 83.7 \\
\midrule
 $DTW_{Chroma}$ & - & - & - & -  & 62.9 & 70.5  & 76.3 & 82.4 \\
\midrule 
 $MLP_{Semigram}$\cite{izmirli2010understanding} & 63.8 & 69.5 & 77.2 & 83.4  & - & -  & - & - \\
\midrule 
 \begin{math}\mathit{SCNN_{STFT}}\end{math}& 65.6 & 71.9 & 78.1 & 84.8  & 67.2 & 73.4  & 78.7 & 85.6 \\
\midrule 
 \begin{math}\mathit{SCNN_{CQT}}\end{math} & 66.4 & 73.1 & 78.7 & 85.3 & 68.1 & 74.8 & 80.1 & 86.7\\
\midrule 
 \begin{math}\mathit{SCNN_{Chroma}}\end{math} & 67.1 & 74.6 & 79.2 & 86.1  & 69.4 & 75.1 & 80.7 & 87.2\\
\midrule 
 \begin{math}\mathit{SCNN_{Sal}}\end{math} & 68.2 & 75.3 & \textbf{81.4} & \textbf{87.8} & 70.3 & 76.7  & 82.1 & 88.4 \\
\midrule 
 \begin{math}\mathit{SCNN_{CQT+DA}}\end{math} & 67.9 & 74.4 & 80.8 & 86.7  & 69.6 & 75.4 & 81.6 & 87.9\\
\midrule 
 \begin{math}\mathit{SCNN_{Sal+DA}}\end{math} & \textbf{69.4} & \textbf{76.4} & 81.2 & 87.5  & \textbf{71.7} & \textbf{78.2} & \textbf{83.3} & \textbf{90.1} \\
\midrule 
\bottomrule
\end{tabular}
\label{results2}
\end{table*}
\vspace{-0.2cm}
 Our Siamese model has four convolutional layers of varying dimensionality followed by a fully connected layer to generate the similarity output. The outputs of each layer are passed through rectified linear units in order to add non-linearity, followed by batch normalization before being passed as inputs to the next layer. The detailed architecture of our model is given in Table \ref{tab:arch}.



We conduct experiments using two different mechanisms for computing the similarity matrix $S_m$:
\begin{itemize}
    \item Using binary labels: We directly employ the outputs of the Siamese CNN, whereby 0 and 1 correspond to similar and dissimilar pairs respectively.
    \item Using distances: We employ the distance $D_W$ computed as part of the loss, which directly corresponds to the dissimilarity between the two inputs.
\end{itemize}
We generate an alignment path through this matrix using DTW, through a readily available implementation in Python \cite{giorgino2009computing}.
For our Siamese models trained without data augmentation, the naming convention we employ is \begin{math} \mathit{SCNN_{x}}\end{math}, where $x$ is the feature representation used during training.
%
We also report results obtained using data augmentation. We generate 20\% additional training samples by employing a random pitch shift of up to \SI{\pm 30} cents, using librosa \cite{mcfee2015librosa}. These models are named \begin{math} \mathit{SCNN_{CQT+DA}}\end{math} and \begin{math} \mathit{SCNN_{Sal+DA}}\end{math} for the CQT  and the salience representations respectively.
\subsection{Results and Discussion}\label{results}


We compare the performance of our models with \emph{MATCH} \cite{dixon2005match}; a DTW algorithm using Chroma features \cite{bartsch2005audio}; and the Multi-Layer Perceptron Model proposed by \cite{izmirli2010understanding} ($MLP_{Semigram}$). We compute the error $e_i$ = ${t_i}^e$ - ${t_i}^r$, defined as the time difference between the alignment positions of corresponding events in the reference ${t_i}^r$ and the estimated alignment time ${t_i}^e$ for score event $i$. We show results for accuracy in percentage for events which are aligned within an error of up to 25 ms, 50ms, 100ms and 200ms respectively. The results obtained by our models are given in Table \ref{results2}. 
Our models outperform DTW-based algorithms that employ handcrafted features as well as an MLP framework which learns binary similarity labels (Table \ref{results2}, rows 1-5). The CQT representation (\begin{math}\mathit{SCNN_{CQT}}\end{math}) yields better results than the STFT representation (\begin{math}\mathit{SCNN_{STFT}}\end{math}), we argue that this  is due to the nature of the CQT, which is a more musically meaningful representation.
Our Siamese model trained using the Chroma representation (\begin{math}\mathit{SCNN_{Chroma}}\end{math}) outperforms the DTW-based method using the same representation ($DTW_{Chroma}$), suggesting that frame similarity learnt from real data is effective at generating robust alignment.
Additionally, we observe the trend that the models trained using a non-binary distance matrix outperform those trained on binary matrices (Table \ref{results2}, columns 6-9). We speculate that thresholding the similarity into binary labels discards potentially useful information and the distances facilitate the DTW algorithm to take better long-term decisions. Both salience representations (\begin{math}\mathit{SCNN_{Sal}}\end{math}) and data augmentation (\begin{math}\mathit{SCNN_{DA}}\end{math}) prove to be effective to improve the performance of our model over   \begin{math}\mathit{SCNN_{CQT}}\end{math}, with salience representations contributing to greater improvements. We posit that using salience representations makes it easier for the model to learn meaningful features from the input representations, since it emphasizes pitched content. Improvements using data augmentation can be attributed to the fact that 
  pianos are not always tuned to $A = 440 Hz$ in the real world, and often the relative intervals are also not tuned perfectly, hence comparison with MIDI files in such cases might lead to false negatives. Data augmentation ensures that the disparity between our training and test conditions is minimized by simulating more real-world like conditions in our training data.  A combination of distance matrix, salience representations and data augmentation  yields the best results (\begin{math}\mathit{SCNN_{Sal+DA}}\end{math}), as can be seen from Table \ref{results2}, row 8, columns 6-9.
  \par Our results demonstrate that frame similarity learning using Siamese neural networks is a promising method for audio-to-score alignment. The principal advantage of this approach over traditional feature choices (like chroma features or MFCCs) is the ability to learn directly from data, which provides higher relevance and adaptability. Both the Siamese network and the pitch salience network are trainable, and thereby adaptable to real world conditions. We plan to explore domain adaptation of our models in the future. 
  A limitation of our method is that it cannot handle structural changes, since DTW generates a monotonically increasing warping path. This could potentially be mitigated by employing an enhanced DTW framework like jump-DTW \cite{fremerey2010handling} alongside our Siamese model. 
\section{Conclusion and Future Work}\label{conclusion}
We presented a novel method for offline audio-to-score alignment using learned similarities via a Siamese convolutional network architecture. 
We demonstrated that our approach is capable of generating robust alignments for piano music across various acoustic conditions. Our models outperform traditional methods based on Dynamic Time Warping that rely on handcrafted features, as well as a Multi Layer Perceptron model which learns binary similarity between audio frames.
We also demonstrated that salience representations and data augmentation are effective techniques to improve alignment accuracy. In the future we plan to incorporate attention into the convolutional models to aid training and improve performance. We would also like to explore other model architectures and work on learning the features as well as the alignments in a completely end-to-end manner. 

\nocite{agrawal2019hybrid}
\bibliographystyle{IEEEbib}
\bibliography{strings-new}

\end{document}